\documentclass[%
 reprint,
superscriptaddress,
 amsmath,amssymb,
 aps,
 pra,
floatfix,
]{revtex4-1}

\usepackage{amsfonts}
\usepackage{amsmath}
\usepackage{amsthm}
\usepackage{mathrsfs}
\usepackage{amscd}
\usepackage{amssymb}
\usepackage{subfigure}
\usepackage{amsxtra}
\usepackage{dcolumn}% Align table columns on decimal point
\usepackage{bm}           % bold math
\usepackage{bbm}
\usepackage{graphicx}
\usepackage{epstopdf}

\usepackage[colorlinks]{hyperref}

\newcommand{\ra}{\rangle}

\newcommand{\ket}[1]{\ensuremath{\left|#1\right\rangle}}

\newcommand{\ketbra}[2]{\ensuremath{| #1 \rangle \langle #2 |}}
\newcommand{\expect}[3]{\ensuremath{\left\langle #1|#2|#3 \right\rangle}}

\newcommand{\cL}{\mathcal{L}}

\newcommand{\dg}{\dagger}
\newcommand{\da}{\dagger}

\newcommand{\Op}[1]{\hat{#1}}

\newcommand{\osigma}{\Op{\sigma}}
\newcommand{\oL}{\Op{L}}
\newcommand{\oH}{\Op{H}}

\newcommand{\stararrow}{\overset{*}{\rightarrow}}

\newcommand{\tr}{\ensuremath{{\rm tr}}}
\newcommand{\dd}{\mathrm{d}}

\begin{document}

\preprint{APS/123-QED}

\title{Refrigeration beyond weak internal coupling}
\author{Stella Seah}
\affiliation{Department of Physics, National University of Singapore, 2 Science Drive 3, Singapore 117542, Singapore}

\author{Stefan Nimmrichter}
\affiliation{Centre for Quantum Technologies, National University of Singapore, 3 Science Drive 2, Singapore 117543, Singapore}

\author{Valerio Scarani}
\affiliation{Department of Physics, National University of Singapore, 2 Science Drive 3, Singapore 117542, Singapore}
\affiliation{Centre for Quantum Technologies, National University of Singapore, 3 Science Drive 2, Singapore 117543, Singapore}

\date{\today}% It is always \today, today,

\begin{abstract}
We investigate the performance of a three-spin quantum absorption refrigerator using a refined open quantum system model valid across all inter-spin coupling strengths. It describes the transition between previous approximate models for the weak and the ultrastrong coupling limit, and it predicts optimal refrigeration for moderately strong coupling, where both approximations are inaccurate. Two effects impede a more effective cooling: the coupling between the spins no longer reduces to a simple resonant energy exchange (the rotating wave approximation fails), and the interactions with the thermal baths become sensitive to the level splitting, thus opening additional heat channels between the reservoirs. We identify the modified conditions of refrigeration as a function of the inter-spin coupling strength, and we show that, contrary to intuition, a high-temperature work reservoir thwarts refrigeration in the strong coupling regime. 
\end{abstract}
\maketitle

\section{Introduction}\label{sec:intro}

Quantum theory has been closely linked with thermodynamics since its beginnings. Interest in this connection has resurged in relation to quantum information science, mainly along two directions. On one hand, axiomatic and foundational aspects are being revisited in the light of entanglement theory. On the other hand, there is the hope that harnessing quantum effects may enhance the performance of miniature thermal machines. Building on some pioneering works \cite{Scovil1959,alicki1979quantum,kosloff1984quantum}, designs for engines \cite{tonner2005,youssef2010,gilz2013,mari2015quantum,kosloff2016flywheel,alex2017rotor,hardal2017,stella2018,alex2018}, refrigerators \cite{linden2010small,levy2012,mu2017,hofer2018} and batteries \cite{binder2015,Campaioli2017,ferraro2017} have been proposed, with the first proof-of-principle experiments reported recently \cite{rossnagel2016,maslennikov2017,klatzow2017experimental}.

These thermal devices typically consist of a few quantum degrees of freedom interacting with thermal reservoirs and can therefore be described naturally using the formalism of open quantum systems. Preliminary studies borrowed standard tools of open quantum systems, but concerns were soon raised about the validity of such heuristic weak-coupling models \cite{rivas2010,correa2013,levy2014,hofer2017,gonzalez2017}. In order to convince the community that something unphysical may arise, the need for modifications was argued based on heat flows between a hot and a cold subsystem, but never in the context of thermal machines. In this paper, all these contributions are integrated to address the optimal performance of thermal machines operating at high output power.

We consider the simplest model of a refrigerator, consisting of three qubits, each in contact with a reservoir. The main parameter is the coupling strength \textit{among} the qubits. In the strong coupling limit, two effects come into play that should be accounted for. First, the interaction Hamiltonian cannot possibly describe only the resonant coupling, since the rotating-wave approximation breaks down \cite{scala2007,hofferberth2007,irish2007,zueco2009,forn2010,beaudoin2011}. Second, it is no longer the case that each qubit sees only the reservoir with which it is in contact: through the coupling, each qubit starts feeling the effect of the other reservoirs as well \cite{rivas2010,correa2013,levy2014,mitchison2015,hofer2017,gonzalez2017}. Looking even more closely, the naive assumption of a frequency-independent interaction with the reservoirs also proves inadequate, therefore a more realistic spectral density must be considered.

By putting together all these modifications, we show how the simplest quantum refrigerator operates optimally in a regime of intermediate coupling. Specifically, we propose a modified master equation that is valid across all coupling strengths, which can be taken as a blueprint for future studies of multipartite open quantum systems such as thermal machines.

\section{The Model}\label{sec:model}

\subsection{System Hamiltonian}

We study an absorption refrigerator system consisting of three interacting qubits, work $(w)$, hot $(h)$ and cold $(c)$, as heat bodies. Their bare energies are described by $\oH_0=\sum_{j=h,c,w} \hbar\omega_j \osigma^+_j\osigma^-_j$. The interaction Hamiltonian has often been chosen as the resonant coupling \cite{linden2010small,skrzypczyk2011smallest,correa2014quantum,brask2015,barra2017local}
\begin{equation}\label{eq:Hres}
\oH_{\rm res} = \hbar g \left( \osigma^+_{h}\osigma^-_{c}\osigma^-_{w} + \osigma^-_{h}\osigma^+_{c}\osigma^+_{w}\right)
\end{equation} on phenomenological grounds. However, the rotating wave approximation fails once the coupling $g$ becomes comparable to the inherent frequencies of the system \cite{scala2007,hofferberth2007,irish2007,zueco2009,forn2010,beaudoin2011}. Here, we work with a more likely candidate of physical Hamiltonian, whose dynamics effectively reduces to that of $\oH_{\rm res}$ only when the qubit energies are resonant ($\omega_h = \omega_c+\omega_w$) and the coupling is weak ($g\ll\omega_j$). We choose the $XXX$-type Hamiltonian
\begin{equation}\label{eq:Hint}
\oH_{\rm int} = \hbar g \osigma^x_{h}\osigma^x_{c}\osigma^x_{w}
\end{equation}
which has the same structure as the harmonic modes interaction used in the experimental realisation \cite{maslennikov2017} and has been proposed as feasible for some spin systems \cite{reiss2003broadband,pachos2004,bermudez2009}. 

The overall system Hamiltonian is $\oH_s = \oH_0 + \oH_{\rm int}$. Its eigenstates can be written in the computational basis $|xyz\ra = |x_h\ra \otimes  |y_c\ra \otimes |z_w\ra $ as
\begin{eqnarray} \label{eq:eigenstates}
\ket{\epsilon_{0\pm}} &=& \frac{1}{\sqrt{2}}\left(\ket{100}\pm\ket{011}\right), \\
\ket{\epsilon_{1\pm}} &=& \frac{1}{N_1^\pm}\left( \ket{110}+\frac{-\omega_c\pm\sqrt{g^2+\omega_c^2}}{g}\ket{001}\right),  \nonumber\\
\ket{\epsilon_{2\pm}} &=& \frac{1}{N_2^\pm}\left( \ket{101}+\frac{-\omega_w\pm\sqrt{g^2+\omega_w^2}}{g}\ket{010}\right),\nonumber \\
\ket{\epsilon_{3\pm}} &=& \frac{1}{N_3^\pm}\left( \ket{111}+\frac{-\omega_h\pm\sqrt{g^2+\omega_h^2}}{g}\ket{000}\right), \nonumber
\end{eqnarray}
with normalization constants $N^{\pm}_a$. The corresponding eigenvalues are
\begin{eqnarray} \label{eq:eigenvalues}
\epsilon_{0\pm} &=& \omega_h\pm g, \quad \epsilon_{1\pm} = \omega_h\pm \sqrt{g^2+\omega_c^2}, \\
\epsilon_{2\pm} &=& \omega_h\pm \sqrt{g^2+\omega_w^2}, \quad \epsilon_{3\pm} = \omega_h\pm \sqrt{g^2+\omega_h^2}. \nonumber
\end{eqnarray}
In the limit where $g\ll\omega_j$, the rotating wave approximation applies, i.e.~the off-resonant terms in the interaction Hamiltonian \eqref{eq:Hint} can be omitted. Accordingly, an expansion to lowest non-vanishing order in $g/\omega_j$ reduces the eigenstates \eqref{eq:eigenstates} and eigenvalues \eqref{eq:eigenvalues} to those of \eqref{eq:Hres}.

\subsection{System-bath coupling}

The interacting qubits are in contact with separate heat baths at temperatures $T_{j=h,c,w}$ representing the hot, the cold, and the work reservoir of the refrigerator. 
We focus on a working regime of comparable qubit energies and moderate thermal excitations, $\omega_w \sim \omega_c$ and $k_B T_j \sim \hbar\omega_j$. 

To describe the open system dynamics, we consider a \textit{weak} system-bath interaction of the linear form $\hat{V} = \sum_{j=h,c,w} \osigma_j^x \otimes \hat{B}_j $ as in the standard Born-Markov approach. We then perform a partial coarse graining over a timescale $\Delta t$ in between the fast system-bath correlation times $\tau_c$ and the characteristic time scale at which the relevant system dynamics takes place \cite{schaller2008coarse,schaller2009coarse,rivas2012open,cresser2017coarse}. The Lindblad jump operators that will appear in the resulting master equation are the spectral components of the coupling operators $\osigma_x^j$ in the interaction picture,
\begin{eqnarray}\label{eq:FDsigX}
\osigma_j^x (t) &=& e^{i\oH_st/\hbar}\osigma_j^xe^{-i\oH_st/\hbar} \\
&=&\sum_{a,b} e^{i(\epsilon_a-\epsilon_b)t} \expect{\epsilon_a}{\osigma_j^x}{\epsilon_b}\ketbra{\epsilon_a}{\epsilon_b} \nonumber\\
&=& \sum_{n=1}^{4} \oL_n^j e^{i \Omega_n^j t} + h.c. \nonumber
\end{eqnarray}
The jump operators $\oL_n^j$ and $\oL_{-n}^j = (\oL_{n}^j)^\da$ mediate transitions between the eight energy eigenstates $\{ |\epsilon_a \ra \}$ with energies $\hbar\epsilon$ of the system, each corresponding to an energy gap of $\pm \hbar\Omega_n^j$ that is accessible by the $j$th bath. In general, there are eight jump operators $\oL_n^j$ (four excitation and four de-excitation terms) associated with each bath $j=h,c,w$. For instance, the first Lindblad operator for the hot bath reads \begin{eqnarray}
\oL_1^h &=& \frac{1}{\sqrt{2}N_3^+}\left(\frac{-\omega_h+\sqrt{g^2+\omega_h^2}}{g}-1\right) \ketbra{\epsilon_{3+}}{\epsilon_{0-}} \nonumber\\
&&+ \frac{1}{\sqrt{2}N_3^-}\left(1-\frac{\omega_h+\sqrt{g^2+\omega_h^2}}{g}\right) \ketbra{\epsilon_{0+}}{\epsilon_{3-}}\label{eq:jumpOpEx}
\end{eqnarray}
with excitation frequency $\Omega_1^h = \sqrt{g^2 + \omega_h^2}+g$. Its hermitian conjugate describes the corresponding de-excitation. The other Lindblad operators have similar expressions. Instead of giving them, we simply list the excitations induced by each bath for reference, denoting with $\overset{*}{\rightarrow}$ those transitions that become accessible at strong coupling and that would vanish when $g\ll\omega_j$:
\begin{eqnarray}\label{eq:jumpops}
\mathrm{Hot}: \ket{\epsilon_{3-}}&\rightarrow&\ket{\epsilon_{0+}} ,\,\ket{\epsilon_{0-}} \rightarrow \ket{\epsilon_{3+}}\\
\ket{\epsilon_{3-}}&\rightarrow&\ket{\epsilon_{0-}} ,\,\ket{\epsilon_{0+}}\rightarrow\ket{\epsilon_{3+}}\nonumber\\
\ket{\epsilon_{1-}}&\rightarrow&\ket{\epsilon_{2+}},\, \ket{\epsilon_{2-}}\rightarrow\ket{\epsilon_{1+}} \nonumber\\
\ket{\epsilon_{1+}}&\stararrow&\ket{\epsilon_{2+}} ,\,\ket{\epsilon_{2-}}\stararrow\ket{\epsilon_{1-}}\nonumber\\
\mathrm{Work}: \ket{\epsilon_{2-}}&\rightarrow&\ket{\epsilon_{0+}} ,\,\ket{\epsilon_{0-}}\rightarrow\ket{\epsilon_{2+}}\nonumber\\
\ket{\epsilon_{2-}}&\rightarrow&\ket{\epsilon_{0-}},\,\ket{\epsilon_{0+}}\rightarrow\ket{\epsilon_{2+}} \nonumber\\
\ket{\epsilon_{3-}}&\rightarrow&\ket{\epsilon_{1-}},\, \ket{\epsilon_{1+}}\rightarrow\ket{\epsilon_{3+}}\nonumber\\
\ket{\epsilon_{1-}}&\stararrow&\ket{\epsilon_{3+}} ,\,\ket{\epsilon_{3-}}\stararrow\ket{\epsilon_{1+}}\nonumber\\
\mathrm{Cold}: \ket{\epsilon_{1-}}&\rightarrow&\ket{\epsilon_{0+}},\,\ket{\epsilon_{0-}}\rightarrow\ket{\epsilon_{1+}} \nonumber\\
\ket{\epsilon_{1-}}&\rightarrow&\ket{\epsilon_{0-}} ,\,\ket{\epsilon_{0+}}\rightarrow\ket{\epsilon_{1+}}\nonumber\\
\ket{\epsilon_{3-}}&\rightarrow&\ket{\epsilon_{2-}},\, \ket{\epsilon_{2+}}\rightarrow\ket{\epsilon_{3+}}  \nonumber\\
\ket{\epsilon_{2-}}&\stararrow&\ket{\epsilon_{3+}} ,\,\ket{\epsilon_{3-}}\stararrow\ket{\epsilon_{2+}}\nonumber
\end{eqnarray}
Each line is associated to a jump operator that is a linear combination of two degenerate excitations with the same transition energy, but different weights; the first line corresponds to the Lindblad operator \eqref{eq:jumpOpEx}. 

The excitation and de-excitation \emph{rates} for the transitions in \eqref{eq:jumpops} are also determined by the rate matrices $\gamma^j(\Delta t)$ of the baths, which crucially depend on the coarse-graining time $\Delta t$. They are defined as \cite{cresser2017coarse}
\begin{eqnarray}\label{eq:exactgamma}
%\gamma_{mn}^j(\Delta t) &=& \frac{1}{\hbar^2}\int_0^{\Delta t} \dd \tau \left[G_j\left(-\tau\right)e^{i\nu_{mn}^j}+G_j\left(\tau\right)e^{-i\nu_{mn}^j}\right]\nonumber\\
%&\times&\frac{e^{i\Omega_{mn}^j\Delta t}e^{-i\Omega_{mn}^j\tau/2}-e^{i\Omega_{mn}^j\tau/2}}{i\Omega_{mn}\Delta t},
\gamma_{mn}^j(\Delta t) &=& \frac{1}{\hbar^2}\int_0^{\Delta t} \dd \tau \left[G_j\left(-\tau\right)e^{i\nu_{mn}^j \tau}+G_j\left(\tau\right)e^{-i\nu_{mn}^j \tau}\right]\nonumber\\
&\times&\frac{e^{i\Omega_{mn}^j\Delta t}e^{-i\Omega_{mn}^j\tau/2}-e^{i\Omega_{mn}^j\tau/2}}{i\Omega_{mn}\Delta t},
\end{eqnarray}
with $\Omega_{mn}^j = \Omega_{m}^j-\Omega_{n}^j$, $\nu_{mn}^j = \left(\Omega_m^j +\Omega_n^j\right)/2$, and 
\begin{equation}\label{eq:Gtau}
G_j\left(\tau\right) = \int_0^\infty J(\omega)\left[ 2\bar{N}_j(\omega) \cos\omega\tau + e^{-i\omega \tau } \right] \dd\omega .
\end{equation}
Here, $\bar{N}_j(\omega) = \left(e^{\hbar\omega/k_B T_j} -1\right)^{-1}$ and $J(\omega)$ is the spectral density of the bath. Specifically, we consider ohmic baths of the same spectral density, 
$J(\omega) \approx \hbar^2 \chi \omega$, where a single dimensionless parameter $\chi$ sets the frequency-proportional thermalization rates to all three baths.
In the weak-coupling limit, the frequency-dependent spectral density and the thermal occupation can be replaced by a constant, as the $j$-th bath then sees only its locally attached spin at frequency $\omega_j$. For the Born-Markov approximation to be valid, we also require that the typical rates $\chi\omega_j \ll k_BT_j$.

\subsection{Markovian master equation}

Assuming we are in the Markovian limit where the characteristic width of $G_j (\tau)$ is narrow, i.e.~the bath-system correlation time $\tau_c \ll \Delta t, |\Omega_{mn}^j|^{-1}$, we can replace the upper limit of the integral in \eqref{eq:exactgamma} with $\infty$ and neglect $\Omega_{mn}^j\tau$ in the second line. This gives us the following expression for the rate matrix $\gamma^j$,
\begin{eqnarray}\label{eq:gammamnOhm}
\gamma_{mn}^j (\Delta t) &\approx& \chi|\nu_{mn}^j| \left[\bar{N}_j \left(|\nu_{mn}^j| \right) + \Theta(-\nu_{mn}^j)\right] \\
&& \times e^{i\Omega_{mn}^j \Delta t/2}\mathrm{sinc}\, \frac{\Omega_{mn}^j\Delta t}{2}\nonumber,
\end{eqnarray}
with $\Theta(\nu)$ the Heaviside function. 
We then arrive at a master equation of the form,
\begin{eqnarray}\label{eq:partialME}
\frac{\dd\rho}{\dd t} &=& -\frac{i}{\hbar}\left[\oH,\rho \right] + \sum_{j=h,c,w} \cL_j \rho, \\
\cL_j \rho &=& \sum_{m,n}\gamma_{mn}^{j}\left[\oL_m^{j}\rho\oL_n^{\dg j}-\frac{1}{2}\left\{\oL_n^{\dg j}\oL_m^{j},\rho\right\}\right]. \nonumber
\end{eqnarray}
The master equation \eqref{eq:partialME} extends across all coupling strengths $g$, eliminating the need to choose between an approximate \emph{local} or \emph{global} description. Typically, a local master equation is valid at weak coupling, where each bath interacts only with the mode it is in direct contact with; it is recovered from \eqref{eq:partialME} in the limit $g\to 0$. The rate matrix and the eight jump operators per bath then simplify and lead to a cancellation of terms in $\cL_j$ such that the bath coupling reduces to the two local transitions $\sigma^\pm_j$,
\begin{eqnarray}
\cL_j \rho &\to& \chi \omega_j \left[ \bar{N}_j (\omega_j) + 1 \right] \left( \osigma_j^{-} \rho \osigma_j^+ - \frac{1}{2} \left\{ \osigma_j^+ \osigma_j^-, \rho \right\} \right) \nonumber \\
&& + \chi \omega_j \bar{N}_j (\omega_j) \left( \osigma_j^{+} \rho \osigma_j^- - \frac{1}{2} \left\{ \osigma_j^- \osigma_j^+, \rho \right\} \right).
\end{eqnarray}
As $g$ increases, the individual baths will cause transitions of various energies that induce spin flips amongst all three spins.

The presence of the cross-terms $m\neq n$ in \eqref{eq:partialME} preserves coherences between the energy eigenstates.
At ultra-strong coupling, each bath can resolve the energy split induced by the inter-spin coupling and causes only incoherent jumps between the eigenmodes of the composite system. The corresponding global master equation is then obtained from \eqref{eq:partialME} by taking only the diagonal elements of the rate matrices $\gamma^j$ into account.

Ideally, the coarse-graining timescale $\Delta t$ should be chosen as to resolve the energy exchange dynamics between the qubits while averaging over their fast free oscillations. In the limit of weakly interacting qubits, the energy level splitting is well approximated by the resonant coupling term \eqref{eq:Hres}, which singles out the relevant intrinsic timescale $g^{-1}$. This degeneracy is lifted at stronger coupling, and for a valid description of the refrigerator system in both regimes, we employ a smaller coarse-graining time, e.g.~$\Delta t = \max{\omega_j^{-1}}$.

If the rate matrices $\gamma^j \geq 0$, one can diagonalize them and bring the master equation to Lindblad form, which would ensure that the time-evolved state is always positive. This was shown to be the case for the exact expression \eqref{eq:exactgamma} \cite{schaller2008coarse}, regardless of the spectral density used. In the Markov approximation, this remains true under the assumption of a flat spectral density \cite{cresser2017coarse}, which is not appropriate in the strong-coupling scenario. In fact, the ohmic expression \eqref{eq:gammamnOhm} may yield negative eigenvalues that are typically orders of magnitude smaller than the largest positive eigenvalue. These negativities vanish in the high-temperature limit $k_B T_j \gg \hbar \omega_j$. For the temperatures considered here as well as for randomly generated parameters, both the time-evolved and stationary states remain positive. Moreover, these states agree with the ones from the local and global master equation models in the respective limits where $g\ll \omega_j$ and $g\gg\omega_j$. 

\section{Refrigerator Performance}\label{sec:performance}

\begin{figure}
   \centering
\includegraphics[width=\columnwidth]{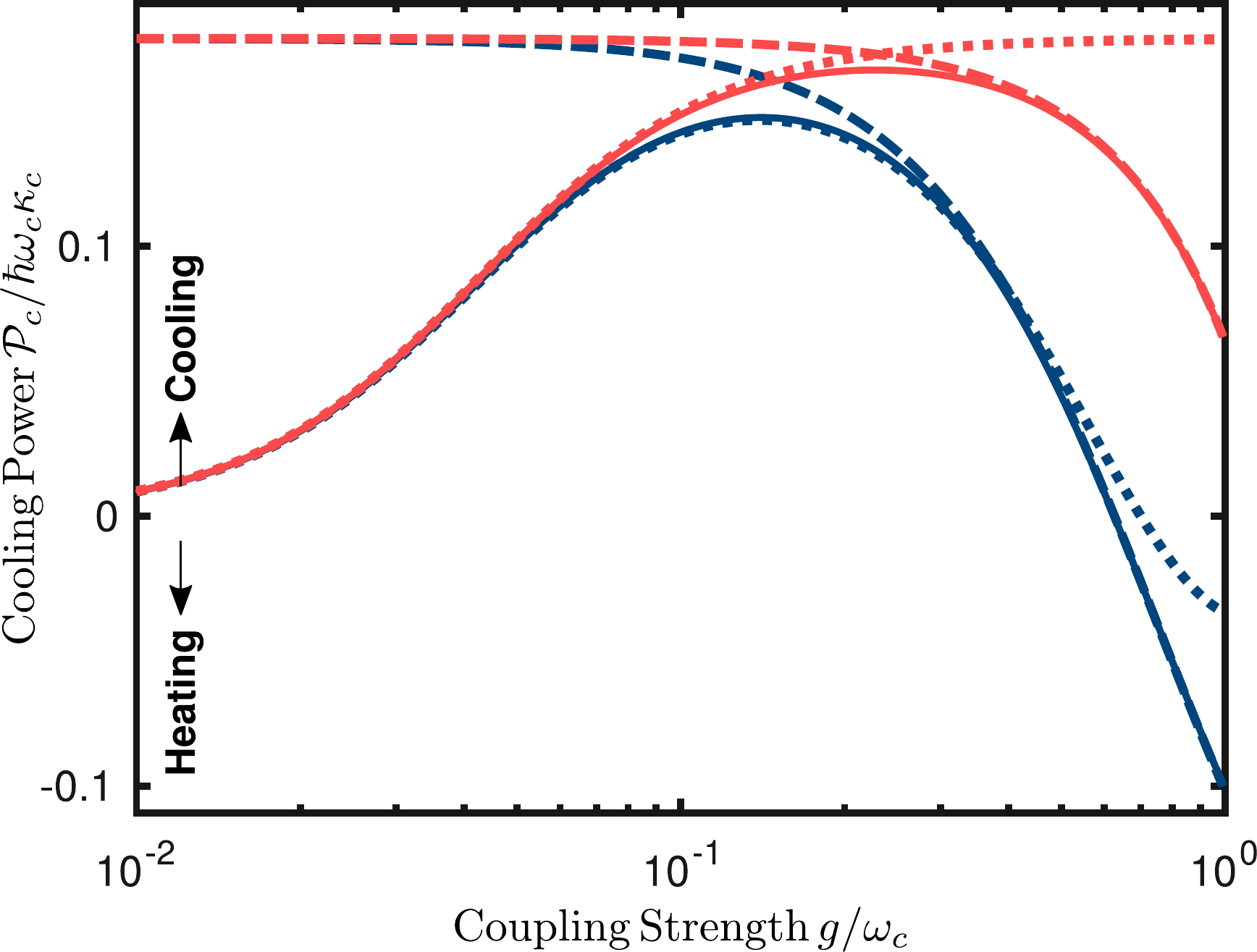}
\caption{\label{fig:compareModels} Cooling power (i.e.~heat flow from the cold bath) $\mathcal{P}_c = \tr (\oH \cL_c \rho_{\infty} )$ as a function of the coupling strength. The solid blue curve is the most accurate prediction within our model, using $\oH_{\rm int}$ and the master equation \eqref{eq:partialME}. The dotted and dashed curves are the predictions obtained using, respectively, the standard local and global master equations. The red family of curves assume the resonant coupling $\oH_{\rm res}$. The qubits have energy gaps $\omega_h = 5\omega_c$ and $\omega_w= 4\omega_c$, and they are in contact with heat baths of temperatures $T_h = 2 T_c$ and $T_w= 8 T_c $, with $T_c= \hbar\omega_c/k_B$ and thermalization constant $\chi = 10^{-2}$.}
\end{figure}

We solve numerically for the steady state $\rho_{\infty}$ of the master equation \eqref{eq:partialME} and compute the average powers $\mathcal{P}_j = \tr (\oH \cL_j \rho_{\infty} )$ associated to the heat flows from the $j$-th bath to the system, where $\mathcal{P}_h + \mathcal{P}_c + \mathcal{P}_w = 0$. In the local weak-coupling limit, which we shall use as a reference, heat enters from each bath in units of $\hbar\omega_j$ at a rate given by $\chi \omega_j$.

\subsection{Cooling power}

Cooling of the cold reservoir occurs when $\mathcal{P}_c > 0$. The cooling power $\mathcal{P}_c$ is plotted in Fig.~\ref{fig:compareModels} against the intrinsic coupling strength $g$ for a fixed thermalization parameter $\chi=10^{-2}$. We observe that the stationary cooling power (blue solid) improves towards a maximum in the strong coupling regime ($g \gtrsim 0.1 \omega_c$), after which it quickly deteriorates and the refrigerator eventually stops cooling at ultra-strong couplings. In subsection \ref{sseff} we shall describe how the maximum cooling power depends on $\chi$.

In the figure, the prediction of our model based on \eqref{eq:Hint} and \eqref{eq:partialME} is compared with those of the local (dotted) and global (dashed) approximations of the master equation, as well as with those of the resonant coupling model (red) using \eqref{eq:Hres}. Our model bridges between the others, agreeing with them in the regimes of their validity. Notice that the assumption of resonant coupling in combination with a local master equation (red dotted) results in a perpetual rise of cooling power with $g$, i.e.~it fails to describe the end of refrigeration at ultra-strong coupling. Conversely, the global approach (dashed) based on a full secular approximation of the bath interactions \cite{correa2013,mitchison2015} is only valid beyond $g/\omega_c> 0.1$ but fails to give the correct behavior as $g \to 0$.

Moreover, while the resonant interaction model (red) agrees well with the $XXX$-coupling at small $g$, it leads to a higher strong-coupling performance. This confirms our expectation that the rotating wave approximation fails there, and it indicates that the off-resonant terms in the interaction Hamiltonian \eqref{eq:Hint} contribute to the performance drop. 
In fact, the rotating wave approximation breaks down \emph{before} the local model, which means that it would be inconsistent to consider a global master equation model in the strong coupling regime and still assume the rotating wave approximation to hold.

%arxiv w/ shading!
\begin{figure}
   \centering
\includegraphics[width=\columnwidth]{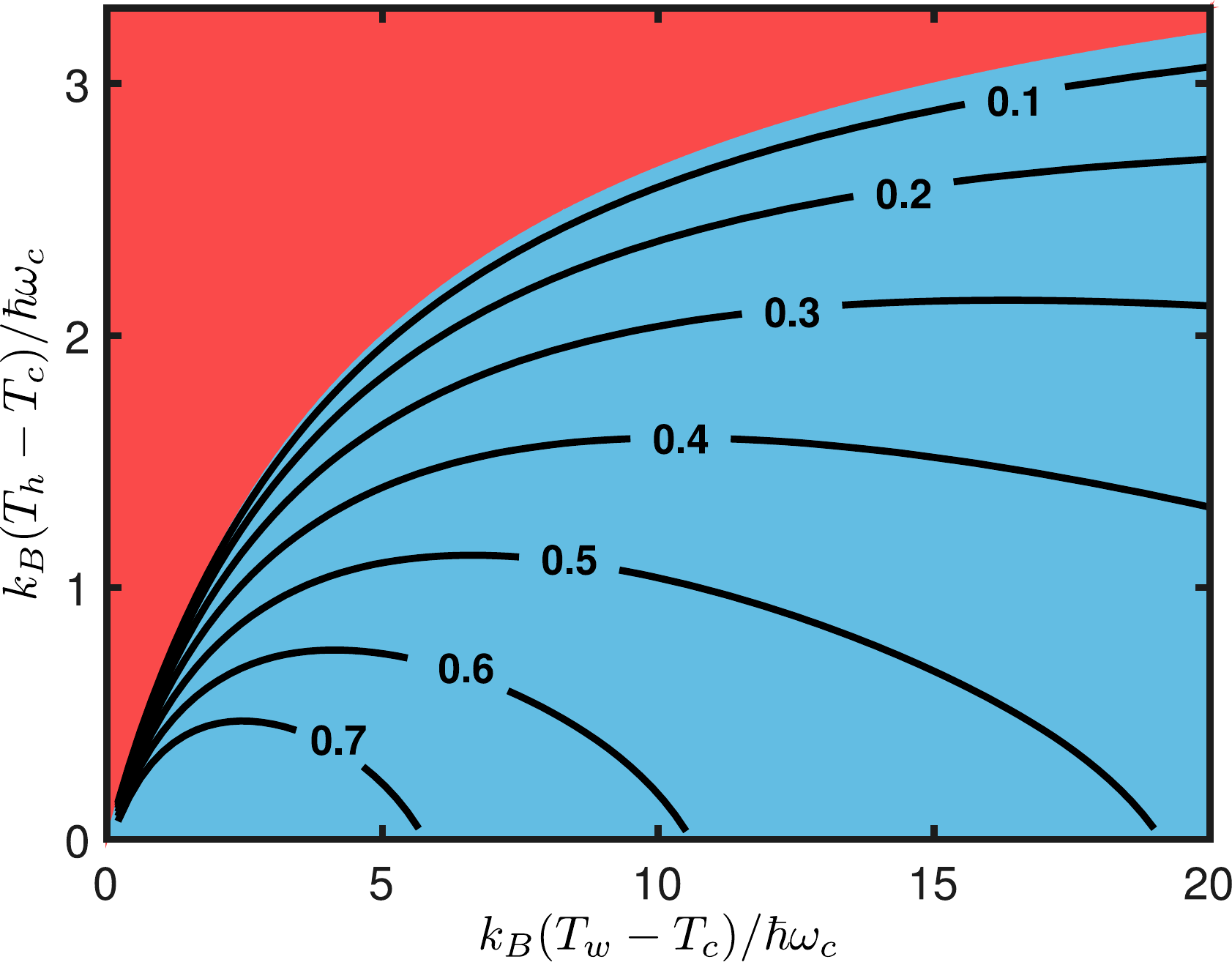}
\caption{\label{fig:coolingwindow} 
Temperature windows of refrigeration for different coupling strengths $g$. 
The shaded region (blue for cooling and red for heating) shows the result of a local weak-coupling model, which is independent of $g$. The solid lines indicate the new boundaries of the windows ($\mathcal{P}_c=0$) for growing values of the coupling strength $g$. The qubit energies are again $\omega_h=5 \omega_c$ and $\omega_w=4 \omega_c$, the cold bath temperature is $k_BT_c = \hbar\omega_c$, and the thermalization parameter is $\chi = 10^{-2}$.}
\end{figure}

\subsection{Cooling window}

Given the cooling behavior observed in Fig.~\ref{fig:compareModels}, we should expect that a strong coupling rate $g$ affects the window of work and hot bath temperatures at which refrigeration occurs. In the weak-coupling limit, where the local resonant model applies, the cooling condition is independent of $g$.  In this case, one can consider the effective temperature $T_v$ of the virtual qubit in the subspace $\left\{\ket{0_h1_w}, \ket{1_h0_w}\right\}$ that swaps excitations with the cold qubit \cite{Brunner2012}. Refrigeration should happen when 
\begin{eqnarray}\label{eq:coolingwin}
T_v = \frac{\omega_c}{\omega_h/T_h-\omega_w/T_w} < T_c,
\end{eqnarray}
which determines the cooling window as a function of $T_w$ and $T_h$. In Fig.~\ref{fig:coolingwindow}, 
it corresponds to the blue-shaded area, and we compare it to the actual temperature bounds for cooling ($\mathcal{P}_c = 0$) at different coupling strengths, ranging from $g/\omega_c = 0.1$ to $0.7$. 
The cooling window shrinks with growing $g$, which we attribute to the fact that, at strong coupling, heat is not just exchanged as intended via the resonant coupling term \eqref{eq:Hres} between the reservoirs, but also through other adversary channels. 

We note here that the master equation \eqref{eq:partialME} would also predict a shrink in cooling window for the resonant interaction \eqref{eq:Hres}, albeit a less significant one. Indeed, the off-resonant interaction terms in \eqref{eq:Hint} open up additional heat exchange channels at strong coupling, which affect the thermally induced transitions \eqref{eq:jumpops} and counteract the desired refrigeration effect. The resonant interaction \eqref{eq:Hres} causes fewer transitions, which leads to a smaller drop in cooling power and would not affect the cooling window as significantly for a given coupling strength $g$.

\begin{figure*}
\centerline{\includegraphics[width=\columnwidth]{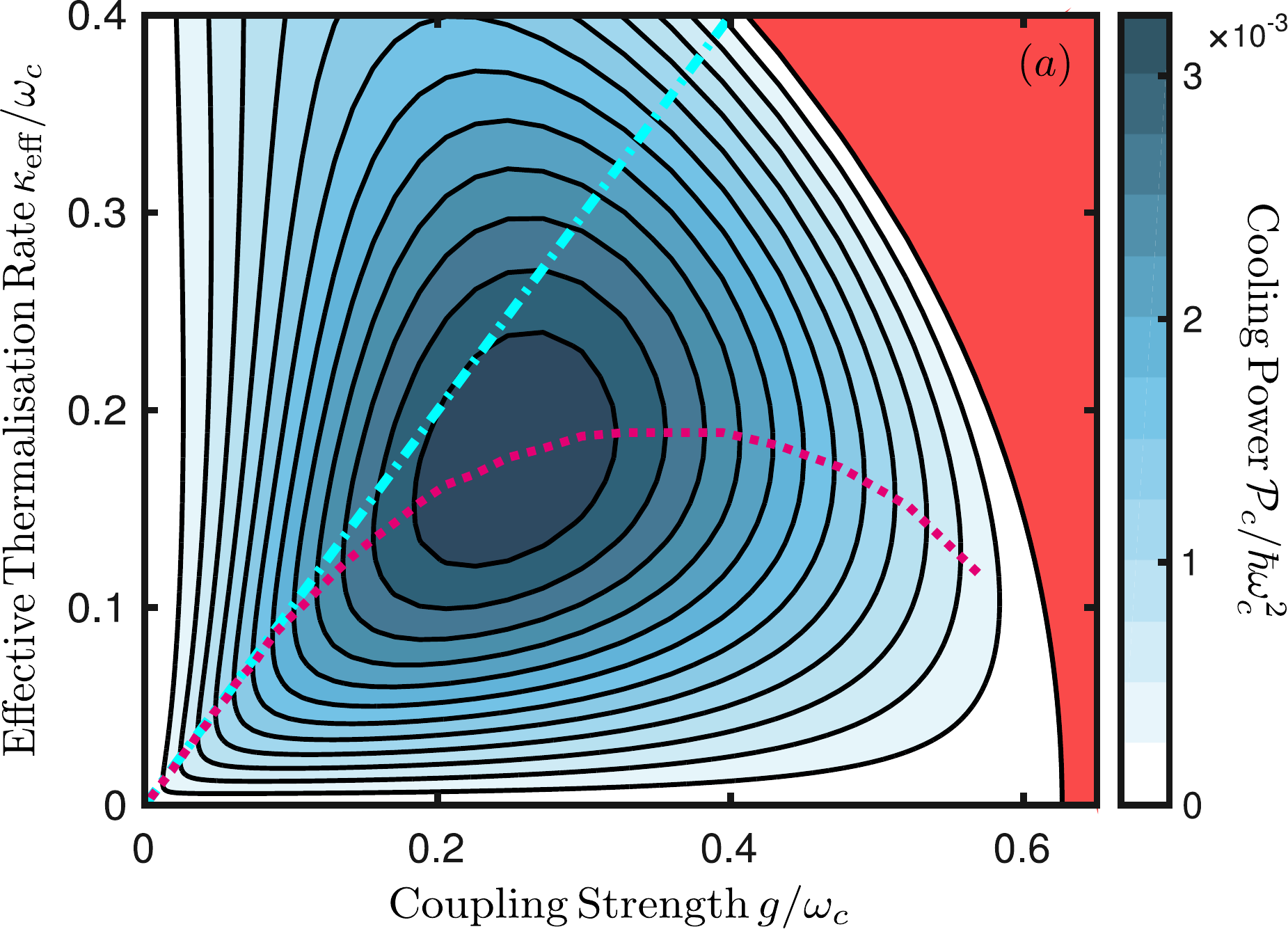}\quad\includegraphics[width=\columnwidth]{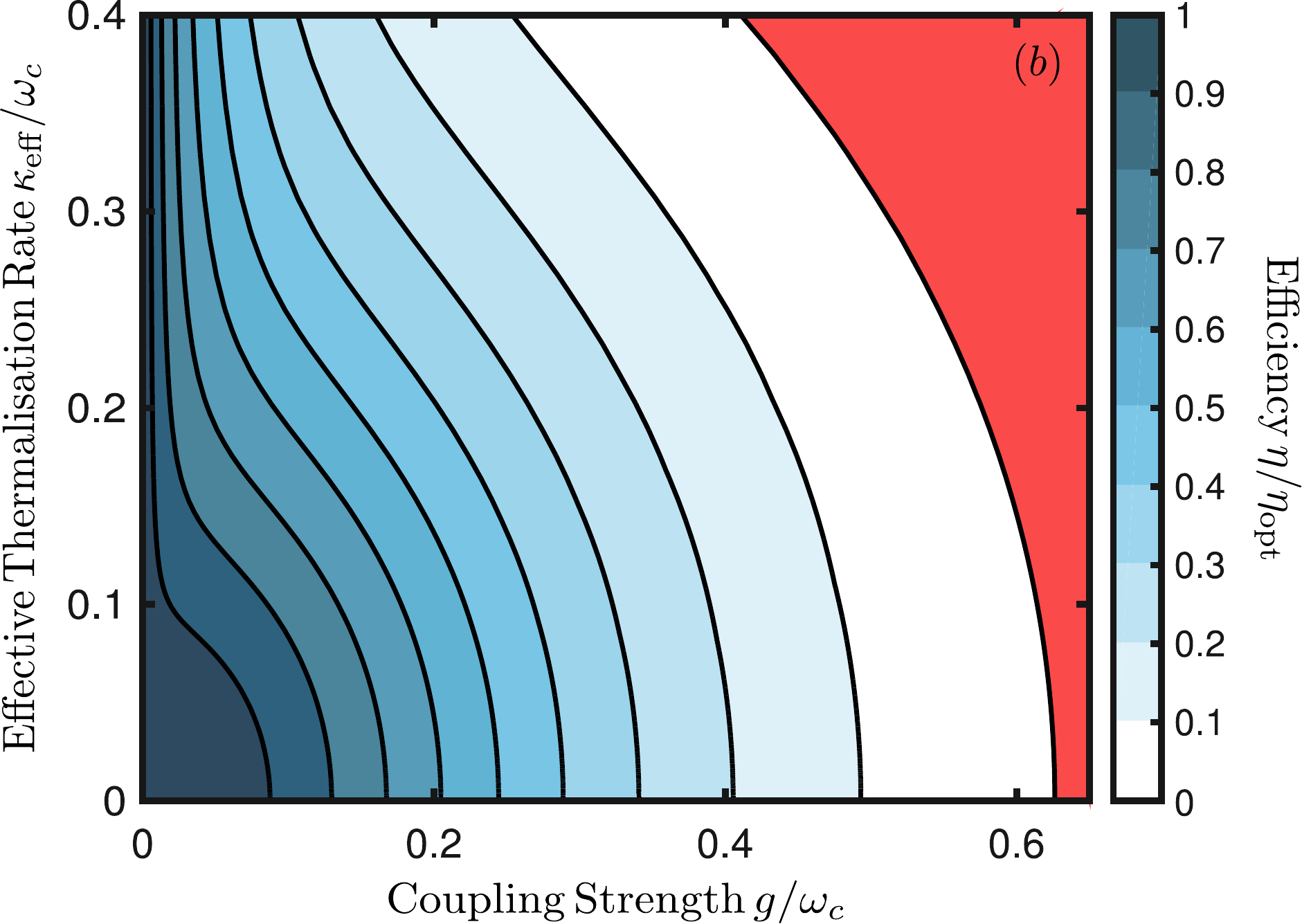}}
\caption{\label{fig:performance}
Refrigerator performance based on (a) cooling power and (b) coefficient of performance for different coupling strengths $g$ and thermalization parameters $\chi$. The latter are given in terms of the effective decoherence rate $\kappa_{\rm eff}$ in \eqref{eq:kappaeff}. The blue dash-dotted line corresponds to $g = \kappa_{\rm{eff}}$, while the red dotted line marks the maximum of cooling power for a given $g$. The qubit energies and temperatures are the same as in Fig.~\ref{fig:compareModels}.}
\end{figure*}

As a result of the off-resonant coupling terms and the interaction-induced level splitting, the heat flows to an extent that high work bath temperatures will ultimately halt refrigerator performance. Indeed, the cooling window for large $g$ does not extend to arbitrarily high $T_w$ and the refrigerator eventually stops working---a counter-intuitive feature if we compare to the local model \eqref{eq:coolingwin}, which always predicts cooling for sufficiently high $T_w$ as long as $T_h/T_c < \omega_h/\omega_c$.

\subsection{Optimal cooling power and efficiency}
\label{sseff}

We now seek to find the optimal refrigerator performance in terms of cooling power $\mathcal{P}_c$ and efficiency $\eta = \mathcal{P}_c/\mathcal{P}_w$ for a given temperature configuration of the external reservoirs. In Fig.~\ref{fig:performance}, we consider a point that lies well within the ideal cooling window \eqref{eq:coolingwin} and show how the performance changes as we vary the coupling strength $g$ and thermalization parameter $\chi$. The blue-shaded contour profile in panel (a) and (b) depict $\mathcal{P}_c$ and $\eta$, respectively, and the red-shaded area indicates where heating instead of cooling occurs. The parameter $\chi$ is expressed in terms of the rate
\begin{equation}\label{eq:kappaeff}
\kappa_{\rm{eff}} = \frac{1}{3}\sum_{j=h,c,w} \chi\omega_j\frac{2\bar{N}_j(\omega_j)+1}{2},
\end{equation}
which, at weak coupling, gives the arithmetric mean of the thermal single-qubit decoherence rates. There we see that the cooling power grows with both the intrinsic coupling rate $g$ and the external rate $\kappa_{\rm eff}$, and the maximum for fixed $g$ is achieved when $\kappa_{\rm eff} \approx g$. We attribute this relation to the characteristic transient dynamics of the refrigerator system, which exhibits an oscillatory enhancement of cooling power at $gt \approx \pi$ before it reaches steady state on the decoherence timescale $\kappa_{\rm eff}^{-1}$ \cite{correa2014quantum,mitchison2015,nimmrichter2017}. The dashed and the dash-dotted lines in (a) represent the points of maximum power at given $g$ and the condition $g=\kappa_{\rm eff}$, respectively. 
To give explicit numbers, the plotted maximum $\mathcal{P}_c/ \hbar \omega_c^2 = 3.2 \times 10^{-3}$ amounts to the refrigerator extracting energy quanta from the cold bath at $1.3\%$ of the coupling rate ($g/\omega_c=0.25$). The local weak-coupling model would predict at best $2\%$, but it is no longer valid here.

\begin{figure*}
\centerline{\includegraphics[width=\textwidth]{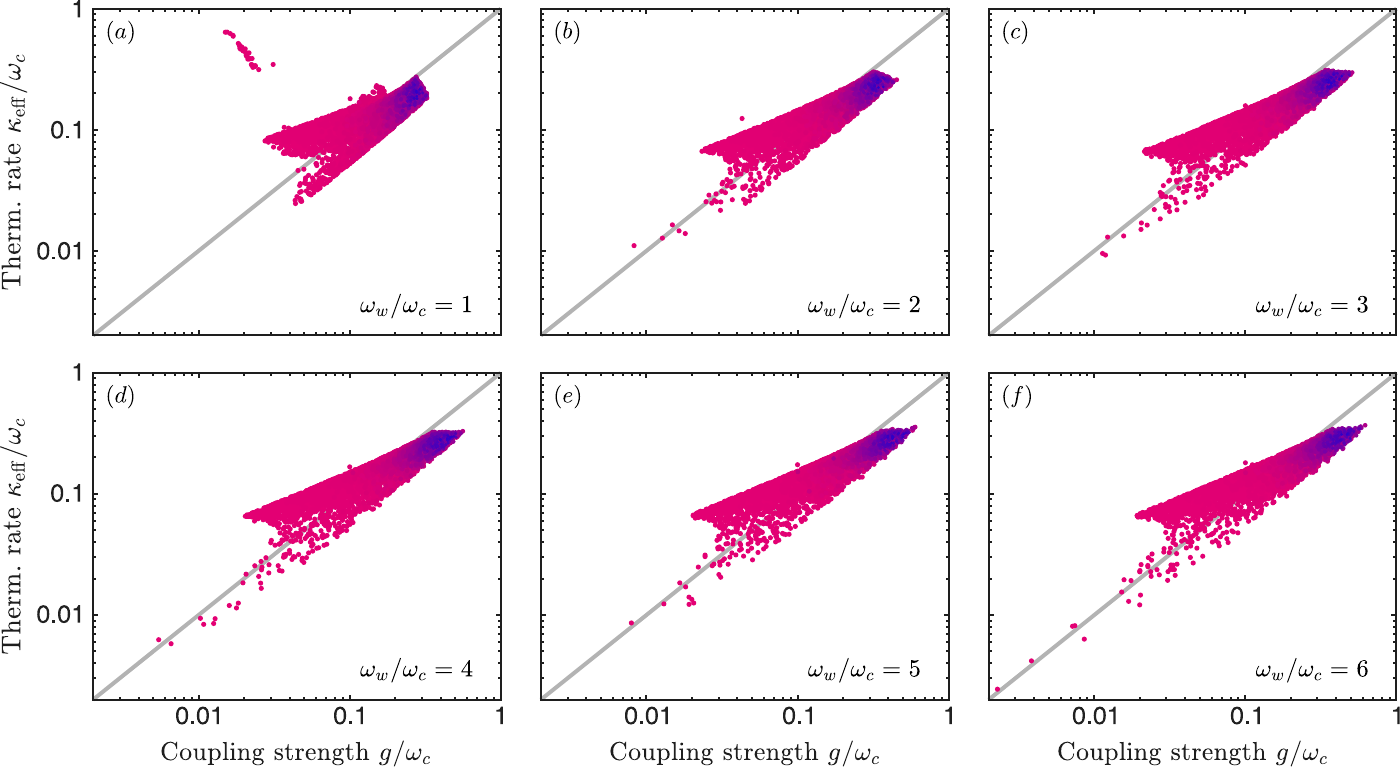}}
\caption{\label{fig:randomT}
Optimal coupling parameters $g$ and $\kappa_{\rm eff}$ for maximum cooling power, given six qubit energies (see $\omega_w/\omega_c$ ratios in panels a-f) and random bath temperatures (5000 values per panel). Each dot corresponds to a positive cooling power, colored according to the magnitude (red to blue). The gray line indicates $g=\kappa_{\rm eff}$. The cold bath temperatures are drawn uniformly from $k_B T_c/\hbar\omega_c \in [1,10]$, then $T_h$ is drawn from $T_h/T_c - 1 \in [0,0.9\omega_w/\omega_c] $ and $T_w$ from $(\omega_h/T_h - \omega_c/T_c)T_w/\omega_w \in [1,10] $, so that cooling is most likely achieved at moderate temperatures.}
\end{figure*}

Thermalization rates greater than $g$ will always thwart refrigeration as they freeze the qubit states close to their initial thermal equilibrium. Note however that the Born-Markov approximation presumes $\chi \ll 1$.
At strong coupling, the maximum moves towards smaller values $\kappa_{\rm eff} \lesssim g$, and these features hold throughout the envisaged working regime with qubit energies and temperatures of similar magnitudes, $\omega_w \gtrsim \omega_c$ and $k_B T_j \sim \hbar \omega_j$. Fig. \ref{fig:randomT} shows the results of this numerical optimization routine for random samples of 5000 different temperatures at six qubit energy ratios. For this we draw random samples of bath temperatures at various qubit energies, and we then vary the coupling strength $g$ and the thermalization parameter $\chi$ to find the maximum steady-state cooling power predicted by \eqref{eq:partialME}.  We restrict to $\chi < 0.1$ to ensure validity of the Born-Markov treatment, and to $g < 1$. Each dot marks the coupling strength $g$ and the parameter $\chi$ (expressed in terms of $\kappa_{\rm eff}$) where a maximum cooling power $\mathcal{P}_c>0$ was obtained for a given set of temperatures $(T_h,T_w,T_c)$. The color indicates the relative value for $\mathcal{P}_c$, from low (red) to high (blue). The gray line corresponds to $\kappa_{\rm eff} = g$, around which the computed maxima congregate. 

The picture changes when we divide the cooling power by the heat inflow from the work reservoir to obtain the refrigerator's coefficient of performance, or efficiency $\eta$, see Fig.~\ref{fig:performance}(b). This assumes its optimal value in the weak coupling limit, where it reduces to a simple expression that is independent of $g$ and $\chi$ and upper-bounded by the Carnot expression as long as cooling occurs \cite{skrzypczyk2011smallest,levy2012},
\begin{equation}
\eta_{\rm opt} = \frac{\omega_c}{\omega_w} \leq \frac{T_c}{T_h - T_c}.
\end{equation}
In fact, this expression hits the Carnot bound in the limit of vanishing cooling power. The efficiency in Fig.~\ref{fig:performance}(b) is plotted in units of this optimal value, which amounts to $25\%$ here. As the couplings increase the efficiency drops, and at the point of maximum cooling power in Fig.~\ref{fig:performance}(a) in particular, it is $9\%$.

Whether at high or low cooling power or efficiency, the stationary state $\rho_{\infty}$ of the refrigerator system does not exhibit any genuine non-classical features across the considered parameter range. 
There are correlations both between the hot and the other two qubits and between the cold and the other two, but we do not observe any entanglement, measured in terms of negativity, for either bipartition. Small negativities do emerge for rather extreme parameter settings, inter alia, a work bath temperature $T_w$ orders of magnitude higher than $T_{h,c}$, in accordance with similar findings \cite{brunner2014entanglement}. 
On the other hand, coherence between the qubits is present, in particular between the states $|100\ra$ and $|011\ra$, but whether this implies genuine quantum behavior depends on how one defines the classical comparison. In fact, for a refrigerator made of harmonic oscillators instead of qubits, coherence-related effects could be reproduced in an entirely classical framework \cite{nimmrichter2017}.

\subsection{Consistency with the second law}

An accurate choice of the underlying master equation model is crucial for a correct thermodynamical description of composite systems, e.g.~to avoid possible violations of the second law \cite{levy2014}, and has therefore been a subject of debate \cite{rivas2010,correa2013,levy2014,hofer2017,gonzalez2017}. Specifically, the second law may or may not hold, depending on whether and how the master equation describes thermalization of the system with each of the reservoirs \cite{alicki1979quantum,levy2014}. 
Here, the only contribution to the steady-state rate of entropy change is associated with the heat flux due to the reservoirs, i.e.~$\dot{\mathcal{S}}=-\sum_j Q_j/T_j$, which grows with $g$ and $\chi$. 

\begin{figure}
   \centering
\includegraphics[width=\columnwidth]{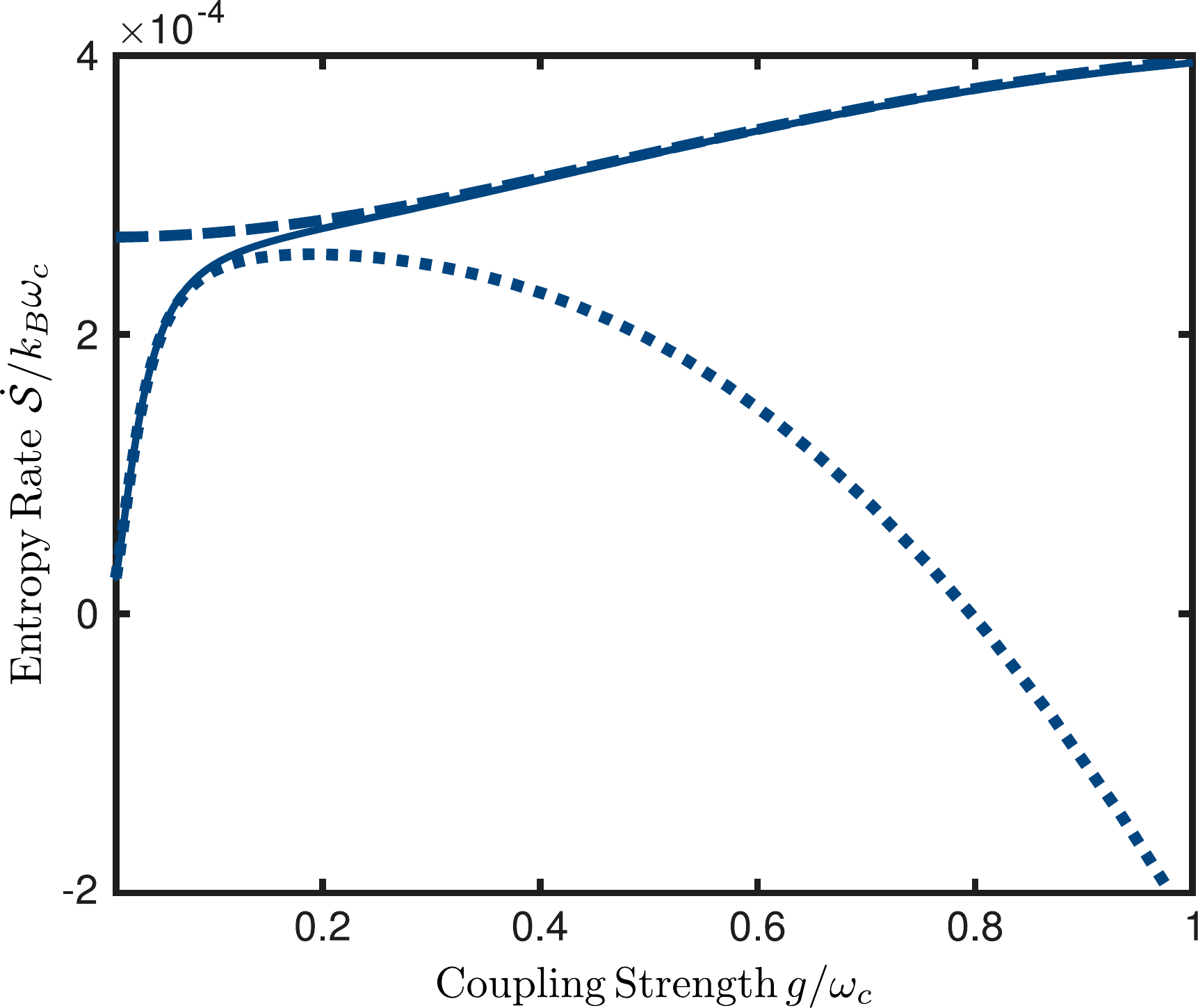}
\caption{\label{fig:violate2law} 
Steady state entropy rates as a function of the coupling strength $g$, using the different master equations, coarse-grained (solid), local (dotted) and global (dashed). Here, the qubits have energy gaps $\omega_h = 5\omega_c$ and $\omega_w= 4\omega_c$, and they are in contact with heat baths of temperatures $T_w = T_h = 2 T_c$, with $T_c = \hbar\omega_c/k_B$ and thermalization constant $\chi = 10^{-2}$. Note that this parameter setting is \emph{not} within the  cooling window specified by \eqref{eq:coolingwin}. }
\end{figure}

We checked that the steady-state entropy rates $\dot{\mathcal{S}}$ predicted by the master equation \eqref{eq:partialME} are positive for all studied parameter settings. 
As an example, we consider $\dot{\mathcal{S}}$ as a function of the coupling strength for $\chi=10^{-2}$ and a fixed temperature setting \emph{outside} the cooling window in Fig.~\ref{fig:violate2law}, comparing the different master equation models.
At large $g$, the local master equation (dotted) would violate the second law, i.e.~predict a heat flow from the cold to the hot reservoir, contrary to the present master equation (solid) and the global approximation (dashed). However, as long as the cooling condition in \eqref{eq:coolingwin} is met, the local resonant model is always compatible with the second law. 

\section{Conclusions}\label{sec:conclusion}

We have characterized the performance of the three-spin absorption refrigerator as a function of the coupling. Our single master equation model mediates between the well-studied weak-coupling approximation, based on a resonant energy exchange in the rotating wave approximation and independent local spin-bath couplings, and a global secular approximation of the combined three-spin system at ultra-strong coupling. With this we could predict that the refrigerator achieves optimal cooling in the intermediate regime of moderately strong couplings $g/\omega_j\sim 0.1$, which thus lends itself to practical implementations, and where neither of the previously studied approximations holds. We also noticed that the rotating wave approximation breaks down before the local bath approximation does, highlighting the role of off-resonant interaction terms at strong coupling that has been overlooked in this context. Additionally, we find that the temperature window for refrigeration shrinks and that hot work reservoirs may become detrimental to (and eventually thwart) cooling with increasing coupling strength. 

\section*{Acknowledgments}

We thank Patrick P.~Hofer and Ronnie Kosloff for helpful discussions. This research is supported by the Singapore Ministry of Education through the Academic Research Fund Tier 3 (Grant No. MOE2012-T3-1-009); and by the same MoE and the National Research Foundation, Prime Minister's Office, Singapore, under the Research Centres of Excellence programme.

%\bibliography{addLit}

\end{document}